\documentclass[twocolumn,superscriptaddress,floatfix,preprintnumbers]{revtex4}
\usepackage[a4paper, total={18cm, 22cm}]{geometry}
\usepackage[utf8]{inputenc}
\usepackage{xcolor}
\usepackage{amsmath}
\usepackage{amssymb}
\usepackage{graphicx}
\usepackage{gensymb}
\usepackage{float}
\usepackage{graphics,color}

\newcommand{\figref}[1]{Fig.~\ref{#1}}

\newcommand{\brk}[1]{\left(#1\right)}

\newcommand{\mm}{\ \text{mm}}

\newcommand{\dn}{\downarrow}
\newcommand{\up}{\uparrow}
\newcommand{\Qdn}{$Q_{\dn}$}
\newcommand{\Qup}{$Q_{\up}$}
\newcommand{\ldn}{$\ell_\dn$}
\newcommand{\lup}{$\ell_\up$}
\newcommand{\Qref}{$Q_{0}$}
\newcommand{\Qtot}{$Q_\textbf{tot}$}

\newcommand{\nturns}{$n_\text{turns}$}
\newcommand{\rhole}{$r_\text{hole}$}

\newcommand{\sfigPrints}{Fig. S1}
\newcommand{\sfigHydraulics}{Fig. S2}
\newcommand{\sfigCalibration}{Fig. S3}
\newcommand{\sfigLeff}{Fig. S4}
\newcommand{\sfigLeffExample}{Fig. S5}

\newcommand{\sfigModel}{Fig. S7}
\newcommand{\sfigDicollapse}{Fig. S8}
\newcommand{\sfigDeformation}{Fig. S9}
\newcommand{\sfigCTScan}{Fig. S10}
\newcommand{\SIdiodicity}{Section 1}
\newcommand{\SImechanical}{Section 2}
\newcommand{\SImodel}{Section 3}
\newcommand{\SIdeformed}{Section 4}
\newcommand{\SIctscan}{Section 5}

\newcommand{\figScheme}{Fig. 2}
\newcommand{\figRigid}{Fig. 3}
\newcommand{\figSoft}{Fig. 4}

\begin{document}

\title{Asymmetric Fluid Flow in Helical Pipes Inspired~by~Shark Intestines}
\author{Ido Levin}
\affiliation{Department of Chemistry, University of Washington – Seattle, Seattle, WA}
\author{Naroa Sadaba}
\affiliation{Department of Chemistry, University of Washington – Seattle, Seattle, WA}
\author{Alshakim Nelson}
\affiliation{Department of Chemistry, University of Washington – Seattle, Seattle, WA}
\author{Sarah L. Keller}
\affiliation{Department of Chemistry, University of Washington – Seattle, Seattle, WA}

        \begin{abstract}
        Unlike human intestines, which are long, hollow tubes, the intestines of sharks and rays contain interior helical structures surrounding a cylindrical hole. One function of these structures may be to create asymmetric flow, favoring passage of fluid down the digestive tract, from anterior to posterior. Here, we design and 3D print biomimetic models of shark intestines, in both rigid and deformable materials. We use the rigid models to test which physical parameters of the interior helices (the pitch, the hole radius, the tilt angle, and the number of turns) yield the largest flow asymmetries. These asymmetries exceed those of traditional Tesla valves, structures specifically designed to create flow asymmetry without any moving parts. When we print the biomimetic models in elastomeric materials so that flow can couple to the structure's shape, flow asymmetry is significantly amplified; it is 7-fold larger in deformable structures than in rigid structures. Last, we 3D-print deformable versions of the intestine of a dogfish shark, based on a tomogram of a biological sample. This biomimic produces flow asymmetry comparable to traditional Tesla valves. The ability to influence the direction of a flow through a structure has applications in biological tissues and artificial devices across many scales, from large industrial pipelines to small microfluidic devices.   
    \end{abstract}
    \maketitle
        
Shark teeth typically elicit more attention than shark intestines (perhaps because the question of whether a shark will eat you elicits more anxiety than the question of whether it will digest you).
Therefore, the beautiful intestinal structures of sharks and rays may come as a surprise.
These structures consist of tubes with interior helices \cite{williams_origin_1972,william_c_hamlett_sharks_1999} that are traditionally thought to confer an advantage by increasing surface area, thereby improving nutrient absorption \cite{chatchavalvanich_histology_2006,honda_morphological_2020}.
Recently, an alternative function was proposed after the discovery of asymmetric flow in shark intestines.
Researchers excised intestines from various shark species and measured flow rates for viscous fluids traveling from anterior to posterior, down the gastrointestinal tract, and in the reverse direction. 
Flow down the tract was faster than the reverse, meaning less peristaltic motion should be needed to push food through the intestines, increasing metabolic efficiency \cite{leigh_shark_2021}. 
This result is notable because asymmetric flow was achieved without the use of flaps like those found in valves of the human heart and stomach.
    
When a fluid conduit imposes asymmetric flow, it behaves like an electrical diode.
The most famous fluidic diodes, \emph{Tesla valves}, were invented over a century ago \cite{tesla_valvular_1920,stith_tesla_2019}.
These quasi-two-dimensional (2D) elements generate vortices and high hydrodynamic drag in only one direction of flow. 
Structures resembling Tesla valves have been found in the lungs of birds \cite{cieri_unidirectional_2016} and have been incorporated into microfluidic circuits \cite{bohm_highly_2022,purwidyantri_tesla_2023}.
The discovery of asymmetric flow in shark intestines is exciting because similar helical structures are potentially scalable for large, 3D applications.

However, it is puzzling how helical shark intestines could operate as Tesla valves. 
The problem lies in the Reynolds number, the ratio between inertial and viscous forces in fluids. The Reynolds number is defined as $\text{Re} \equiv \frac{u L}{\nu}$, where $u$, $L$, and $\nu$ are the flow rate, characteristic length scale, and kinematic viscosity of the fluid, respectively.
A disadvantage of Tesla valves is that flow asymmetry is high only when the Reynolds number is high \cite{nobakht_numerical_2013,thompson_numerical_2014,nguyen_early_2021,liu_scaling_2022}, requiring high flow rates, large length scales, and/or low fluid viscosities. 
In contrast, food flowing through intestines is viscous and flows at low velocity, resulting in $\text{Re} \sim 10^{-4} \relbar 10^{-1}$ \cite{akahashi_flow_2011}.
In this low Reynolds number regime, flow is reversible, and simple Tesla valves are inefficient \cite{purcell_life_1977,brody_biotechnology_1996}.
Mathematically, asymmetric flow through rigid pipes at low Reynolds numbers should be impossible because it violates reciprocity \cite{masoud_reciprocal_2019}. 
However, shark intestines are not rigid; they are soft tissues with mechanical stiffnesses on the order of kiloPascals \cite{stewart_quantitative_2018}.
    
\begin{figure}%[tbhp]
\centering
\includegraphics[width=8.5 cm]{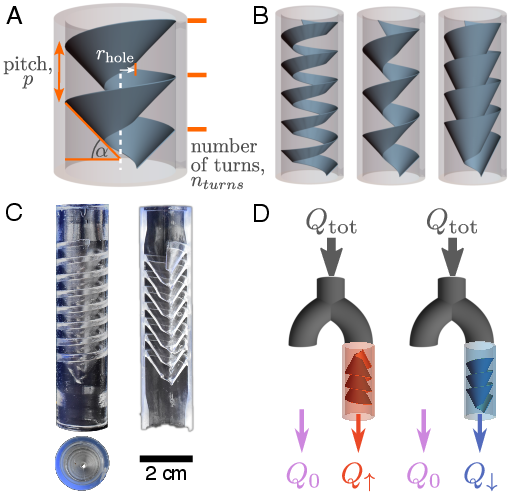}
\caption{
Designing, printing, and evaluating biomimetic helical pipes.
(A)~Pipe with a 10~mm radius, a thick (2~mm) outer wall, and a thin (0.5~mm) inner helical sheet. The sheet begins at radius \rhole\ from the vertical axis.
The interior helix is characterized by the pitch, $p$, the hole radius, \rhole, the tilt angle, $\alpha$, and the number of turns, \nturns.
The tilt angle defines the flow direction: up/down when the imaginary tip of the cone-like profile points upstream/downstream, respectively (shown in panel D).
(B)~Three helical pipe designs. 
(C)~Left: Photo of a helical pipe printed in rigid material: side and top views. Right: 3D print of a transverse section of the same pipe with a view of the inner structure (right).
(D)~The total flow, \Qtot\ (gray), is split between a short reference pipe (\Qref, purple) and the helical pipe in the up (\Qup, red) or down (\Qdn, blue) orientation.
}
\label{fig:design}
\end{figure}

Here, we hypothesize that flow-induced deformation of intestinal structures improves their performance as Tesla valves, enabling them to operate efficiently even at low Reynolds numbers.
In other words, a deformable Tesla valve, whether comprised of biological tissue or elastomeric materials, might generate high flow asymmetry that does not vanish in the limit of low Reynolds number.
To test the effect of deformability on flow asymmetry, we 3D print biomimetic helical pipes of both soft and hard polymeric materials. 
Using well-defined test structures, we measure the influence of the interior helical pitch, hole, angle, and length on flow asymmetry (\figref{fig:design}), and we fit our data to a phenomenological model. 
We then compare flow asymmetry in the test structures with direct replicas of a shark intestine.

\section*{Results}
    
Our experiments address three questions: (1) are tubes with interior helices good candidates for Tesla valves, (2) if so, which features of the design maximize flow asymmetry, and (3) can flow-induced deformation of the structures improve their flow asymmetry?
Here, we designed structures inspired by shark intestines: an outer pipe with a sturdy cylindrical wall (2~mm thick) and an inner, thin helical sheet (0.5~mm thick) that is more likely to bend. 
    
We printed helical pipes from two different materials, a thermoset rigid plastic that minimizes deformability (\figref{fig:design}C) and a soft elastomer that maximizes deformability. Even though the elastomer is among the softest products commercially available for 3D printing (with stiffness of only 1~MPa; see SI \SImechanical), it is much stiffer than shark intestines (with stiffness $\sim$1~kPa) \cite{stewart_quantitative_2018}. Therefore, higher flow rates, $Q$, are required to deform elastomeric helices than to deform intestines. Here, we use flow rates of $Q \approx 100$~cc/s, corresponding to Reynolds numbers $Re \sim 10^4$.
In this regime, flow can be turbulent, so flow asymmetry may arise in rigid pipes as well as in deformable pipes.
Consequently, we use rigid pipes to test the effect of the pipe's geometry, uncoupling it from the mechanical deformation.
Then, we use soft pipes to measure how much the deformability enhances the asymmetry with respect to a rigid pipe of the same design.

\subsection*{Large flow asymmetries in rigid, helical pipes}

The helical pipes in \figref{fig:design} are fully characterized by only a few geometrical parameters. 
These include the pitch ($p$, the vertical ascent with each revolution), the radius of the interior hole (\rhole), the tilt angle ($\alpha$, which breaks up-down symmetry), the number of helical turns (\nturns), and the pipe radius ($R$, from the center to the outside edge). 
We fixed the pipe radius at $10\mm$ and varied the rest of the parameters to generate a large family of structures (\figref{fig:design}B, and \sfigPrints). 
To determine how each parameter influences flow asymmetry, we first established a measurement method.

\begin{figure*}
\centering
\includegraphics[width=17.8 cm]{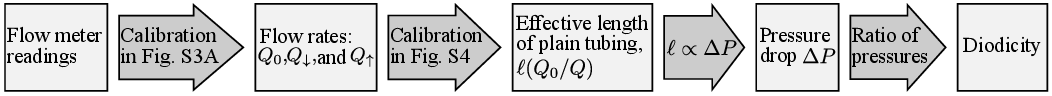}
\caption{
Schematic of diodicity measurement.
Flow meter values are translated into flow rates by the calibration curve in \sfigCalibration A.
Next, ratios of flow rates through both branches of the apparatus are translated into an effective length of hollow tubing, $\ell$, using the calibration curve in \sfigLeff.
This effective length is proportional to the pressure drop.
The diodicity is the ratio between the pressure drop through the helical pipe in the oriented in the up direction to a helical pipe in the down orientation. 
}
\label{fig:scheme}
\end{figure*}

Asymmetric fluid flow is benchmarked by "diodicity," $Di$ \cite{nobakht_numerical_2013,thompson_numerical_2014}:
\begin{equation}
    Di (Q) \equiv \frac{\Delta P_\uparrow (Q)}{\Delta P_\downarrow (Q)},
    \label{eq:diodicity}
\end{equation}
where $\Delta P_\downarrow$ and $\Delta P_\uparrow$ are pressure drops measured across a device in its forward and reverse directions, respectively.
Since it is harder to push fluid through a Tesla valve in the reverse direction, $Di\ge1$, where $Di=1$ indicates no asymmetry.
Here, our device is a helical pipe, and we measure flow rates rather than pressure drops (\figref{fig:design}D). The cone of the helix is oriented either \emph{down} (the forward direction) or \emph{up} (the reverse direction, \figref{fig:design}D).
We convert flow in the pipe, $Q$, into equivalent lengths of hollow tubes, $\ell$, which correspond to pressure drops, $\Delta P$ (\figref{fig:scheme} and Materials and Methods).

\begin{figure}
\centering
\includegraphics[width=8.7 cm]{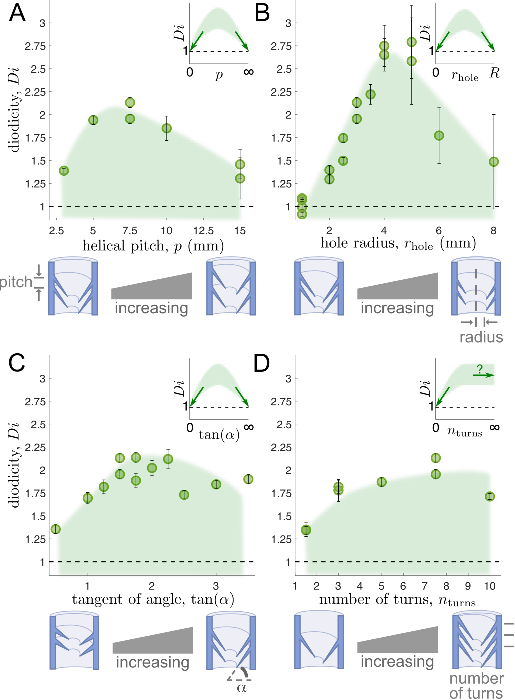}
\caption{
\emph{Rigid} helical pipes result in large flow asymmetries. 
Effect of four parameters on flow diodicity; schematics illustrate each parameter.
(A)~Effect of pitch, $p$, on diodicity for pipes with \rhole~=~3~mm, $\tan(\alpha)$ =~1.5, and \nturns~=~7.5.
(B)~Effect of hole radius, \rhole, on diodicity for pipes with $p$~=~7.5~mm, $\tan(\alpha)$ = 1.5, and \nturns~=~7.5.
(C)~Effect of tilt angle, $\alpha$, on diodicity for pipes with $p$~=~7.5~mm, \rhole~=~3~mm, and \nturns~=~7.5.
(D)~Effect of number of turns, \nturns, on diodicity for pipes with $p$~=~7.5~mm, \rhole~=~3~mm, and $\tan(\alpha)$ = 1.5.
Shaded regions on the graphs are only to guide the eye.
Limiting cases for each parameter are illustrated in the insets.}
\label{fig:rigid}
\end{figure}

Limiting cases yield a qualitative understanding of how diodicity should vary with the four experimental parameters for helical pipes ($p$, \rhole\, $\alpha$, \nturns). These limits appear in the insets of \figref{fig:rigid}.

\begin{enumerate}
    \item Pitch, $p$: 
    As $p$~$\rightarrow$~0, the helical pipe resembles a cylindrical tube with thick walls, so $Di~\rightarrow$~1.
    At the other extreme, as $p\rightarrow \infty$, its effect on the flow diminishes, so $Di~\rightarrow$~1.
    
    \item Hole radius, \rhole:
    As \rhole $\rightarrow$~0, the helical pipe resembles a coiled, hollow tube, for which no flow asymmetry is expected, so $Di~\rightarrow 1$. 
    At the other extreme, as \rhole\ approaches the radius of the pipe, the pipe resembles a cylindrical tube. Therefore, as \rhole$\rightarrow$~10~cm, $Di~\rightarrow$~1.
    
    \item Angle, $\alpha$:
    As $\alpha$ $\rightarrow$~0, the helical pipe becomes symmetric. Therefore, as $\tan(\alpha)\rightarrow 0$, $Di \rightarrow 1$.
    As the angle approaches 90$^\circ$, the helical pipe resembles a cylindrical tube. Therefore, as $\tan(\alpha)\rightarrow \infty$, $Di~\rightarrow$~1.

    \item Number of turns, \nturns:
    As \nturns $\rightarrow$~0, the inner helix vanishes, leaving a cylindrical tube, so $Di~\rightarrow 1$.
    As \nturns~increases, asymmetry is expected to increase, but its detailed behavior is unknown.
    For some designs of Tesla valves, diodicity approaches an asymptote as the number of identical elements in series increases~\cite{thompson_numerical_2014,wang_diodicity_2023}.
\end{enumerate}

These limiting cases imply that flow asymmetry is maximized at intermediate values of pitch, \rhole, and angle, and reaches an asymptotic value as the number of turns increases. Because our system operates far from the regime of laminar flow, exact values of diodicities are difficult to predict. Therefore, we go beyond qualitative limits to quantitatively measure diodicities in rigid helical pipes. 

Our most striking result for these rigid helical pipes is that nearly all values of the parameters tested induce large flow asymmetries(\figref{fig:rigid}).
The diodicity values we measure (in many cases, $2\le Di \le 3)$ are large compared to diodicities measured in traditional Tesla valves.
The literature contains only a few experimentally measured Tesla valve diodicities, with reported values of $Di\sim2$ \cite{nguyen_early_2021,bohm_highly_2022}.
In contrast, many numerical analyses have been conducted on Tesla valve designs; nearly all yield values of $1<Di<2$ \cite{nobakht_numerical_2013,thompson_numerical_2014,bohm_highly_2022}.
In one extreme case, a numerical shape optimization involving an intensive survey of designs and parameters found a highly efficient single-valve design with $Di\sim2-4$ \cite{liu_scaling_2022}.

In other words, biomimetic helical pipes (even rigid structures without flow-induced deformation) can perform better than most Tesla valves, comparable to heavily optimized designs. 
In \figref{fig:rigid}, data in all four panels represent perturbations from an initial parameter set of $p$~=~15~mm, \rhole~=~3~mm, $\tan(\alpha)$~=~1.5, and \nturns~=~7.5. 
When we subsequently vary pitch and \rhole, clear maxima appear at $p$~=~$\sim$7.5~mm and at \rhole~=~4-5~mm. 
As the number of turns increases, diodicity appears to plateau, consistent with asymptotic diodicities in microfluidic Tesla valves \cite{thompson_numerical_2014}.

To understand how the design parameters affect diodicity within the range of values in \figref{fig:rigid}, we developed a phenomenological model. First, we reduced the dimensionality of the problem by defining a helical length, $h\equiv\sqrt{p^2+\pi^2 r_\text{hole}^2}$, where diodicity peaks at intermediate values of the dimensionless variable $h/R$ (\sfigDicollapse).
Then, we tested a model that the effective length of a helical tube, $\ell$, depends separately on three parameters: $h$, $\alpha$, and \nturns, where $\ell = H(h)\times A(\alpha)\times N(n_\text{turns})$. We found good agreement between this model and the data over most parameter ranges in our experiments (SI \SImodel\ and \sfigModel).

\subsection*{Huge flow asymmetry in deformable, helical pipes}
    
\begin{figure}
\centering
\includegraphics[width=8.7 cm]{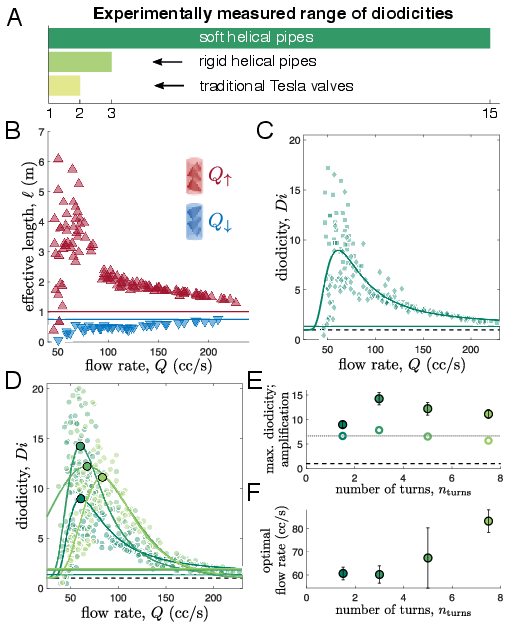}
\caption{
\emph{Deformable} helical pipes result in high diodicities that vary with flow rate. 
(A)~Ranges of measured diodicities of traditional Tesla valves \cite{nguyen_early_2021} and of our rigid and soft helical pipes.
(B)~Effective lengths of a deformable pipe ($p$~=~7.5~mm, \rhole~=~3~mm, $\tan(\alpha)$~=~1.5, and \nturns~=~1.5) connected in the up (red) and down (blue) flow direction. 
Values for an equivalent rigid pipe are shown as red ($\ell~=~$1.0~m) and blue ($\ell~=~$0.75~m) horizontal lines.
(C)~Diodicity as a function of flow rate through the pipe in panel B (squares), through a deformable pipe with identical parameters (diamonds), and through a rigid pipe (horizontal green line).
Scatter in the data of the two deformable pipes is indistinguishable.
An optimal flow rate that yields the highest diodicity is estimated by fitting a skewed Gaussian (thick green line).
(D)~Skewed Gaussian fits (solid lines) of diodicity versus flow rate for deformable pipes with different numbers of turns.
Large circles show the maximum diodicity for each fit.
(E)~Maximum diodicities from panel~D (filled circles) and amplification of diodicity (unfilled circles, the maximum diodicity of the deformable pipe divided by the diodicity of the equivalent rigid pipe). The dotted line shows the average, 6.7-fold amplification.
(F)~Optimal flow rates from panel~D. At the highest number of turns, the highest flow rate is required to reach the maximum diodicity.}
\label{fig:soft}
\end{figure}

Deformation of helical pipes printed from soft elastomers \emph{amplifies} flow asymmetry with respect to nearly all rigid pipes of the same design (\figref{fig:soft}). In the absence of a theoretical model, we fit diodicities in \figref{fig:soft} to skewed Gaussians. The peaks of these Gaussians correspond to diodicities of $\sim$10-15, a flow asymmetry that is roughly 7-fold higher than in rigid helical pipes or in traditional Tesla valves (\figref{fig:soft}A,E). 

Deformable helical pipes do not generate a fixed value of diodicity, unlike rigid versions. Instead, $Di$ is a function of the flow rate, $Q$. This contrast between rigid and deformable pipes is analogous to the observation that an electrical resistor has a fixed resistance, whereas a diode does not.
Flow rates corresponding to the highest diodicities increase with the number of turns of the helical pipe (\figref{fig:soft}F).
This makes sense: larger flow rates are needed to deform a larger number of helical blades. In contrast, the degree to which deformable pipes amplify diodicity relative to rigid pipes is independent of the number of turns. A separate observation is that higher diodicities are accompanied by higher scatter in the data, independent of whether helical pipes have the same parameters  (\figref{fig:soft}C) or different numbers of helical turns (\figref{fig:soft}D). 

The amplification of diodicity in deformable pipes relative to rigid pipes could arise from changes in the flow rate for helices oriented in the up direction, \Qup, helices in the down direction, \Qdn, or both. Here, we find that nearly all amplification arises from changes in \Qup, resulting in high values of the effective length \lup, whereas \ldn\ is roughly constant (\figref{fig:soft}B). This large change in \Qup\ implies that the interior blades of helical pipes deform more when the helix is oriented up. In contrast, if any deformation occurs when helical pipes are oriented down, it appears to slightly facilitate flow, resulting in lower values of \ldn\ than in rigid pipes (\figref{fig:soft}B).

To test that deformation of the inner helical sheet leads to higher diodicity, we simulated the deformation of a single-turn helix in analog scenarios of uniform load and gravity (see SI \SIdeformed\ and \sfigDeformation). We find that most of the deformation is localized to the top and bottom edge of the helix (rather than uniformly changing a parameter such as the angle). Then, we 3D-printed the deformed configurations in rigid materials, locking in the deformation, and we measured flow through the resulting stiff structures. As in deformable pipes, rigid pipes with locked-in deformations produce large effective lengths when the helix is oriented upward, whereas deformations produce little change in effective length when the helix is oriented downward. As a result, the diodicity of the single-turn helical pipe with locked-in deformations is almost 3-fold larger than of the undeformed pipe.

\subsection*{Large flow asymmetries in rigid analogs of shark intestines}

The well-parameterized helical pipes in \figref{fig:design} were inspired by more complex structures in the intestines of sharks and rays. These structures vary from species to species \cite{chatchavalvanich_histology_2006,honda_morphological_2020,leigh_shark_2021}. Here, we created 3D models of the spiral intestine of a dogfish shark (\emph{Centroscyllium nigrum}), based on digital tomograms \cite{leigh_shark_2021}. We 3D-printed rigid versions of the model at three different lengths: the full model, the top two-thirds, and the top third (\sfigCTScan). All three versions generated flow asymmetries in the range of $\sim1.15-1.4$, comparable with traditional Tesla valves (see SI \SIctscan\ and \sfigCTScan). Of course, sharks are not rigid. In vivo ultrasound reveals twisting, contraction waves, and undulatory waves in shark intestines \cite{tomita_narrowing_2023}. A fuller understanding of flow asymmetries in these structures will require ultrasoft materials coupled to biomimetic motions.

\section*{Discussion and Summary}

This research began by posing three questions: Will helical pipes act as Tesla valves? What parameters might increase flow asymmetry? And will it be enhanced by flow-induced deformations of the inner helical structure?
We found that rigid helical pipes impose high flow asymmetries, exceeding those of most traditional Tesla valves (and on par with highly optimized Tesla valves). We are unaware of any other experimentally measured structures that achieve such high flow asymmetries in the absence of moving parts. Optimization of the pipe's design parameters would likely push the flow asymmetries to even higher values. An added bonus is that the pipes are three-dimensional, so they have the potential to accommodate larger fluid volumes than traditional quasi-two-dimensional Tesla valves, and they may find applications in larger commercial devices.

When we reproduced the helical pipes in deformable materials, we found very large flow asymmetries, roughly 7-fold higher than in rigid pipes or Tesla valves. Typically, interactions between elastic structures and fluids are considered from the standpoint of how changes in the structure’s shape affect flow in the surrounding fluid, resulting in mobility, as in Taylor swimmers \cite{taylor_analysis_1997,lauga_hydrodynamics_2009}. Another common vantage is to consider how hydraulic flow induces large changes in a structure’s shape, as in soft robots \cite{wehner_integrated_2016,drotman_electronics-free_2021}. In our system, the viewpoint is different: fluid flow is affected by the deformation it imposes on the elastic structure. 

In our experiments, we used one of the softest commercially available elastomers for 3D printing. Given that the field of 3D printing is quickly evolving \cite{narupai_100th_2020}, softer materials like hydrogels may soon be widely available \cite{chimene_hydrogel_2020,aldana_trends_2021}. However, an ongoing challenge is finding very soft materials that can withstand high deformations \cite{sedlacik_race_2021}. As softer elastomeric materials are developed and integrated into helical pipe designs, we would expect flow asymmetries to arise at lower Reynolds numbers.

\section*{Materials and Methods}
    \subsection*{3D printing helical pipes}
    3D models of helical pipes with different values of pitch, hole radius, angle, and number of turns were generated using \emph{Wolfram Mathematica 13.1} and 
    are available in STL format in \href{https://doi.org/10.5061/dryad.4j0zpc8mt}{Dryad}.
    An adaptor (2~cm long and 4~mm thick) was added to the end of each pipe to connect it to the flow apparatus.
    
    Models were printed at high spatial resolution and accuracy on a stereolithographic (SLA) 3D printer (\figref{fig:design}C and \sfigPrints).
    Specifically, models were 3D printed vertically on a Formlabs \emph{Form2} printer with a layer thickness of 0.1~mm, using \emph{Clear V4} (Formlabs) resin for rigid pipes and \emph{Elastic 50A} (Formlabs; named for its hardness rating of 50 Shore A) for deformable pipes.
    Printed helical pipes were washed several times in isopropyl alcohol and dried in ambient conditions for a few hours.
    Finally, each end of the printed pipe was joined to short PVC tubing (9.5~mm inner diameter, 2.5~cm length). The junction was sealed with epoxy on the interior surfaces and silicone sealant on the outside.
    
    \subsection*{Measuring diodicity}
    In brief, we split the total fluid flow, \Qtot, into two branches with the same resistance (\sfigCalibration B). Each branch contained a flow meter (\sfigCalibration A). We then affixed a helical pipe to one branch, while keeping the other branch empty, as a reference. We measured the flow rate through the reference pipe, \Qref, and the flow rate through the helical pipe, which is either \Qup\ or down \Qdn, depending on whether the helix was oriented up or down. Each helical pipe had a flow resistance equivalent to the resistance through a long span of tubing. The length, $\ell$, of this tubing was proportional to the pressure drop $\Delta P$.
    
    In more detail, schematics of both branches of the flow apparatus are given in  \figref{fig:design}D and \sfigHydraulics, and the method of converting flow rates to pressure drops is given in \figref{fig:scheme} and the Supplemental Information.
    Both branches were equipped with a flow meter (DIGITEN G1/2" Water Flow Hall Sensor 1-30~L/min FL-408) in series with flexible PVC tubing (12.7~mm inner diameter, 30~cm length).
    Each flow meter was previously calibrated to maximize accuracy, as in \sfigCalibration A.
    In each experiment, a helical pipe was attached to the right branch of the flow apparatus, and the left branch remained empty.
    Total flow was controlled by the output of a standard water faucet, and values of \Qref\ (left arm) and \Qdn/\Qup\ (right arm) were recorded by hand.
    Flows through each branch corresponded to an effective length of tubing (\sfigLeff), which in turn corresponded to a pressure drop.
    The ratio between the pressure drops in both pipe orientations was the diodicity (see SI \SIdiodicity for more information). 
    For (only) rigid pipes, \Qref\ depended linearly on both \Qdn\ and \Qup, with negligible offset.
    Therefore, we fit the flow in both directions with linear models,with no intercept terms (\sfigLeffExample) and used the slopes to estimate $Di$ (which is independent of $Q$).

\section*{Data availability}
    All measured data and 3D models are available on Dryad, doi:10.5061/dryad.4j0zpc8mt.

\section*{Acknowledgement}
    I.L. was supported by the Washington Research Foundation and by the Fulbright Foundation. S.L.K. acknowledges funding from NSF MCB-1925731 and MCB-2325819, and A.N. acknowledges funding from NSF EFMA-2223537.
    The authors are grateful to Samantha Leigh for providing shark tomogram data.

% Bibliography
\bibliography{shark_intestines}

%%%%%%%%%% Merge with supplemental materials %%%%%%%%%%
\pagebreak
\widetext
\begin{center}
\textbf{\large Supplemental Materials for\\Asymmetric Fluid Flow in Helical Pipes Inspired~by~Shark Intestines}
\end{center}
%%%%%%%%%% Merge with supplemental materials %%%%%%%%%%
%%%%%%%%%% Prefix a "S" to all equations, figures, tables and reset the counter %%%%%%%%%%
\setcounter{equation}{0}
\setcounter{figure}{0}
\setcounter{table}{0}
\setcounter{page}{1}
\makeatletter
\renewcommand{\theequation}{S\arabic{equation}}
\renewcommand{\thefigure}{S\arabic{figure}}
\renewcommand{\bibnumfmt}[1]{[S#1]}
\renewcommand{\citenumfont}[1]{S#1}
%%%%%%%%%% Prefix a "S" to all equations, figures, tables and reset the counter %%%%%%%%%%

\section{Diodicity measurement}
\begin{figure}
\centering
\includegraphics[width=10 cm]{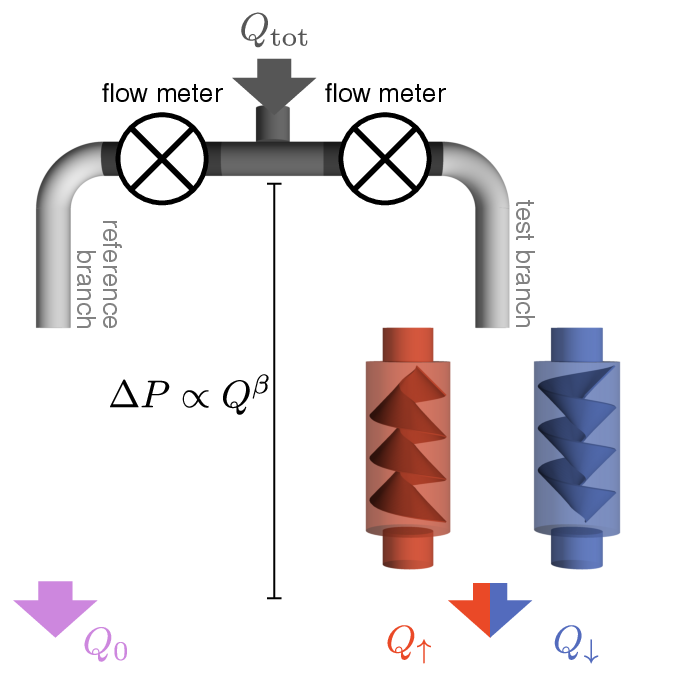}
\caption{
Schematic of the parallel flow apparatus.
An input flow, \Qtot, is split into two branches, each of which passes through a previously-calibrated flow meter.
In each experiment, a helical tube is connected to the test branch (right), while the reference branch (left) remains unchanged. In the first half of each experiment, the helical tube is connected in the ``down" direction, with the tip of the cone pointing downstream.
As the input flow, \Qtot, is varied, flow rates through the left branch (\Qref) and right branch (\Qdn) are recorded.
In the second half of the experiment, the helical tube is connected in the ``up" direction, which yields a corresponding data set for \Qref\ and \Qup.
The flow rates \Qref, \Qdn\ and \Qdn\ are converted to pressure drops via a phenomenological model: $\Delta P \propto Q^\beta$, where $\beta$ is fit from experimental data (Supporting Information Text, Section 1)}.
\label{sfig:hydraulics}
\end{figure}

\begin{figure}
\centering
\includegraphics[width=11.4 cm]{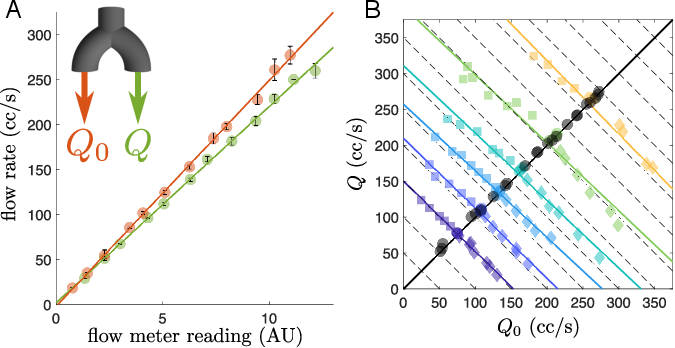}
\caption{
Calibration of flow meters.
(A) The flow apparatus splits into reference (left) and test (right) branches, each of which contains a flow meter (see \figref{sfig:hydraulics}). 
The reference and test flow meters were separately calibrated by measuring the rate at which a container with volume markings filled ($Q_0$ and $Q$, respectively) and by plotting that rate against the reading on the flow meter (in arbitrary units, AU).
The resulting linear fits were used in all experiments to convert the meter reading to the actual flow rate.
In subsequent experiments, samples are attached to the test branch, and the reference branch remains empty. 
(B) The symmetry of the flow apparatus is verified by recording flow rates when both branches are empty (black points).
A linear fit (black line), shows asymmetry of 2\% (slope of 0.98).
As a test of the flow apparatus, each branch is pinched to increase its hydrodynamic resistance for several values of the total flow (colored points).
Once a branch is pinched, the flow through the other branch increases, conserving the total flow, as expected (conservation of flow produces data parallel to the gray dashed lines.)
}
\label{sfig:calibration}
\end{figure}

This section elaborates on the diodicity measurement scheme summarized in \figScheme.
Each branch of our apparatus is fit with a flow meter (\figref{sfig:hydraulics}).
The reading from each meter is converted to a flow rate by a calibration curve to improve its accuracy (\figref{sfig:calibration}).
The variable $Q$ corresponds to the flow rate (in cc/s) through the test branch, to which samples are attached, and \Qref\ corresponds to the reference branch, which remains empty.

These values are converted to a pressure drop, using the constitutive relation of our fluidic apparatus.
In general, the form this relation takes depends on the Reynolds number of the system and the behavior of fluid flow through the apparatus.
The pressure drop in a system with laminar flow behaves according to Ohm's Law: $\Delta P = R_1 Q$, where $R_1$ is the hydraulic resistance.
The pressure drop in a system with fully turbulent flow is often described by a quadratic relation for: $\Delta P = R_2 Q^2$.
Our experiments fall in an intermediate range of Reynolds numbers, which requires us to fit a phenomenological model $\Delta P = R Q^\beta$, where we expect $\beta$ to be between 1 and 2.
In the limits of both laminar flow and turbulent flow, the hydraulic resistance is proportional to the length of the pipe, $L$, such that $R \propto L$. This relation should hold for intermediate values of the Reynolds number as well.

\begin{figure}
\centering
\includegraphics[width=11.4 cm]{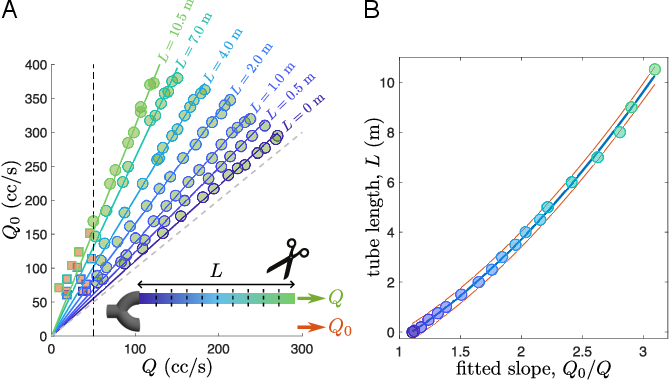}
\caption{
Converting flow rate, $Q$, into pressure drop, $\Delta P$, via a phenomenological model.
(A) An input flow, \Qtot, is split into two branches. A long piece of hollow tubing (initial length~$L=$~10.5~m, inner diameter~=~9.5~mm) is connected to the test branch, while the reference branch remains unchanged. 
As the input flow is varied, flow rates through the reference branch (\Qref) and test branch ($Q$) are recorded.
Then the tube is cut to a slightly shorter length and the experiment is repeated. This process is iterated until $L$~=~0.
Data for \Qref\ versus $Q$ at each length are fit to straight lines passing through the origin.
Data for low flow rates ($Q<$~50~cc/s) are excluded (filled squares) because they do not follow the linear trend and because this regime is excluded in the analysis of the helical pipes.
(B) Fitting a phenomenological power law model to the slopes in Panel A ($L=A (Q_0/Q)^\beta + c$).
The fitted exponent, $\beta = 1.75 \pm 0.15$ agrees well with the data (blue line), and has narrow 95\% confidence intervals (red lines).
The color of the symbols corresponds to the length of the pipe.
}
\label{sfig:Leff}
\end{figure}

As the hydraulic resistance of the tested pipe (in the test branch of the apparatus) \emph{increases}, the ratio $Q_0/Q$ \emph{increases}.
We utilize this fact to find the phenomenological exponent $\beta$ by testing tubular pipes of varying lengths (\figref{sfig:Leff}).
We vary the total flow rate, and we record the flow rates through both branches.
We repeat this procedure for many different lengths of pipe, and for each length we fit a linear model to extract the ratio $Q_0/Q$.
We then use these values to fit a phenomenological power law, $L=A (Q_0/Q)^\beta + c$ and find $\beta = 1.75 \pm 0.15$.

Next, for each pipe tested, we convert the pair of values ${Q,Q_0}$ into an effective length, $\ell (Q/Q_0)$.
We repeat this conversion for both orientations of each helical pipe.
Since, $\Delta P \propto \ell$, the diodicity is given by the ratio between effective lengths: 
\[
    Di \equiv \frac{\Delta P_\uparrow}{\Delta P_\downarrow} = \frac{\ell (Q_\uparrow/Q)}{\ell (Q_\downarrow/Q)}
\]
Because the effective lengths of the rigid pipes are constant, we first fit a linear model to $Q$ and \Qref, and only then convert the fitted slope to the effective length.
The resulting effective lengths vary smoothly with the various parameters (\figref{sfig:Leff_rigid}).

\section{Mechanical properties of our materials}
We tested the mechanical properties of both materials using an Instron 5585H load frame with a 2 kN load cell, at a speed of 5~mm/min until mechanical failure of the sample. Standard "dog-bone test specimen" shapes were 3D-printed in the Form2 SLA printer, following ISO 527-2 for the rigid material and ISO 37 for soft material.
We treated the test samples in the same way as the helical pipes: we washed them with isopropanol and then water until the samples were no longer sticky. 
The results of the tests are summarized in the table below:
\begin{center}
    \begin{tabular}{||c || c | c||} 
        \hline
        Material & Rigid (\emph{Clear V4}) &  Soft (\emph{Elastic 50A})\\
        \hline\hline
        Elastic modulus (MPa) & $555\pm 76$ & $1.18\pm 0.06 $ \\ 
        \hline
        Tensile stress (MPa) & $30.13\pm 0.32 $  & $0.03\pm 0.002 $ \\ 
        \hline
        Tensile strain (\%) & $41\pm 0.3 $ & $200\pm 1 $ \\ 
        \hline
    \end{tabular}
    \label{tab:moduli}
\end{center}

\section{Phenomenological model for rigid, helical pipes}
\begin{figure}
\centering
\includegraphics[width=11.4 cm]{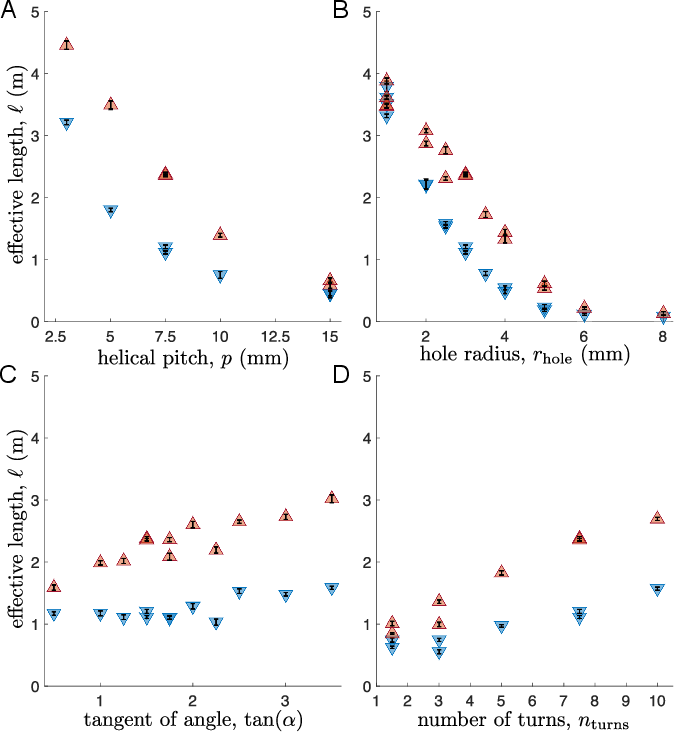}
\caption{
Effect of four parameters on the effective length, $\ell$, of helical pipes oriented in the down direction (blue, downward triangles) and in the up direction (red, upward triangles).
(A)~Effect of the pitch, $p$, on the effective length for pipes with \rhole~=~3~mm, $\tan(\alpha)$~=~1.5, and \nturns~=~7.5.
(B)~Effect of the hole radius, \rhole, on the effective length for pipes with $p$~=~7.5~mm, $\tan(\alpha)$~=~1.5, and \nturns~=~7.5.
(C)~Effect of the tilt angle, $\alpha$, on the effective length for pipes with $p$~=~7.5~mm, \rhole~=~3~mm, and \nturns~=~7.5.
(D)~Effect of the number of turns, \nturns, on the effective length for pipes with $p$~=~7.5~mm, \rhole~=~3~mm, and $\tan(\alpha)$~=~1.5.
}
\label{sfig:Leff_rigid}
\end{figure}

\begin{figure}
\centering
\includegraphics[width=17.5 cm]{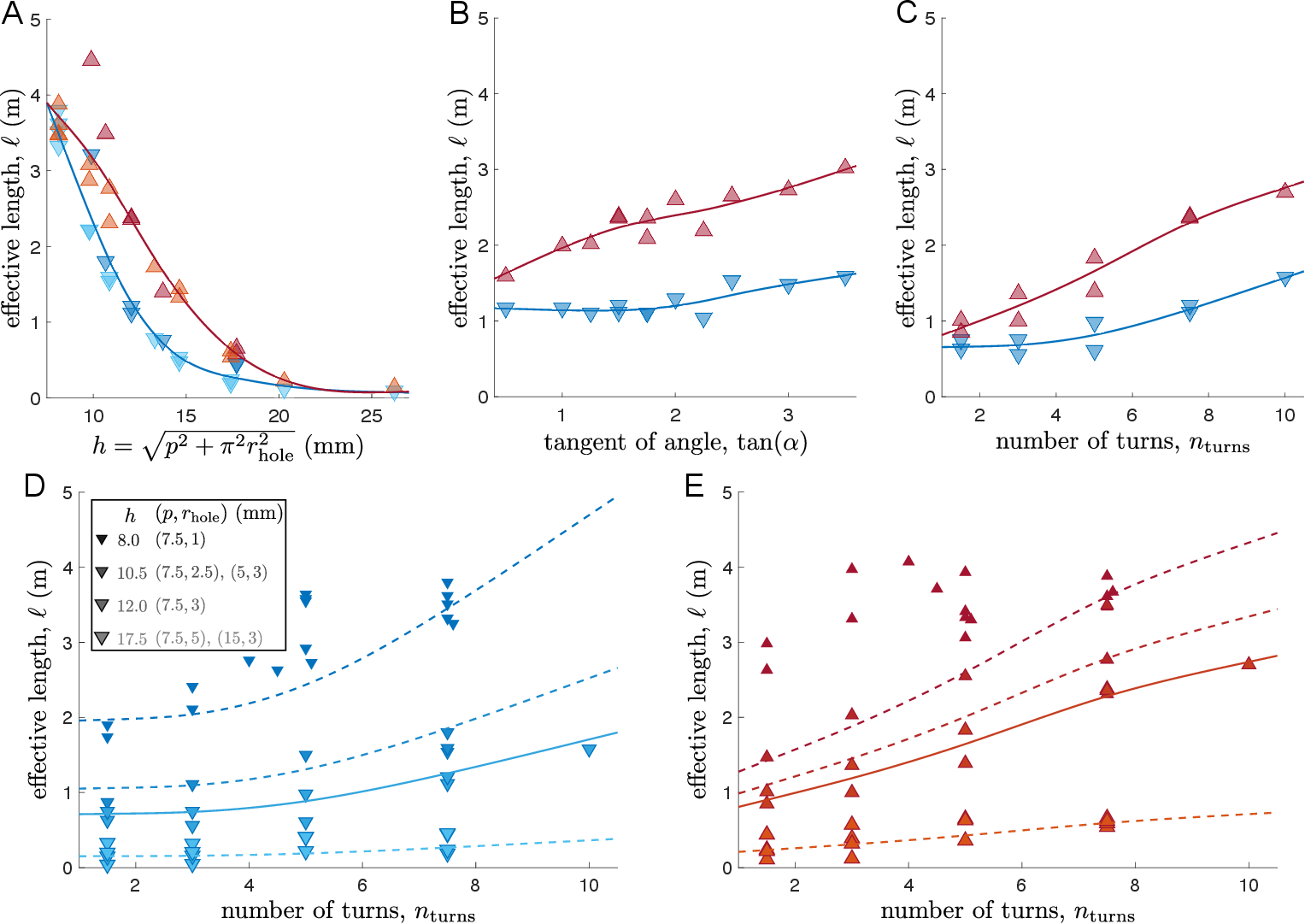}
\caption{
Phenomenological model for the effective length, $\ell$, of helical pipes.
(A)~By introducing a new variable, helical length $h\equiv\sqrt{p^2+\pi^2 r_\text{hole}^2}$, data for effective length vs. pitch $p$ (darker symbols, \figref{sfig:Leff_rigid}A) and for effective length vs. hole radius \rhole\ (lighter symbols, \figref{sfig:Leff_rigid}B) collapse. Data for flow through helical pipes oriented with their cones in the up (upstream) direction is shown in red, and the down (downstream) direction is shown in blue. Data for each orientation are fit with smoothed spline functions, denoted $H(h)$.
(B)~Data for effective length vs. the tangent of the angle are fit with smoothed spline functions denoted $A(\tan(\alpha))$, using data from \figref{sfig:Leff_rigid}C at $h \approx 12.0$~mm. (C)~Data for effective length vs. the number of helical turns are fit with smoothed spline functions denoted $N(n_\text{turns})$, using data from \figref{sfig:Leff_rigid}D for $h \approx 12.0$~mm.
(D) and (E) The data and fits (solid lines) for $h \approx 12.0$~mm are repeated from Panel C. The solid line is a fit of the model $\ell=H(h)\times A(\alpha)\times N(n_\text{turns})$ and differs only slightly from the fit in C due to normalization of the product $H\times A\times N$. The dashed lines are outputs from the model $\ell=H(h)\times A(\alpha)\times N(n_\text{turns})$, applied at $h \approx 8$~mm, $h \approx 10.5$~mm, and $h \approx 17.5$~mm, with no fitting parameters. To test the model, data corresponding to the three new values of $h$ (from five new datasets for rigid pipes with varying $p$ and \rhole\ values) are overlain on the plots. Agreement between the model and the data are good for all data for helical pipes in the down direction (Panel D) and for data at larger values of $h$ in the up direction (Panel E). Poor agreement between the model and data at $h \approx 8$~mm in Panel E is expected from the large scatter in the corresponding data in Panel A and in \figref{sfig:Leff_rigid}A,B.
}
\label{sfig:model}
\end{figure}

We use the helical symmetry of the rigid pipes to introduce a phenomenological model that reduces the number of parameters required to fit the effective length, $\ell$, of helical pipes.
First, we define a parameter "helical length", $h\equiv\sqrt{p^2+\pi^2 r_\text{hole}^2}$.
The helical length is the path length swept out by one turn of a helix with pitch, $p$, and radius, $r_\text{hole}/2$.
The helical length parameter collapses data for $\ell$ vs. $p$ (\figref{sfig:Leff_rigid}A) and data for $\ell$ vs. \rhole\ (\figref{sfig:Leff_rigid}B) onto the curves in \figref{sfig:model}A, which are fit by splines.

\begin{figure}
\centering
\includegraphics[width=17.5 cm]{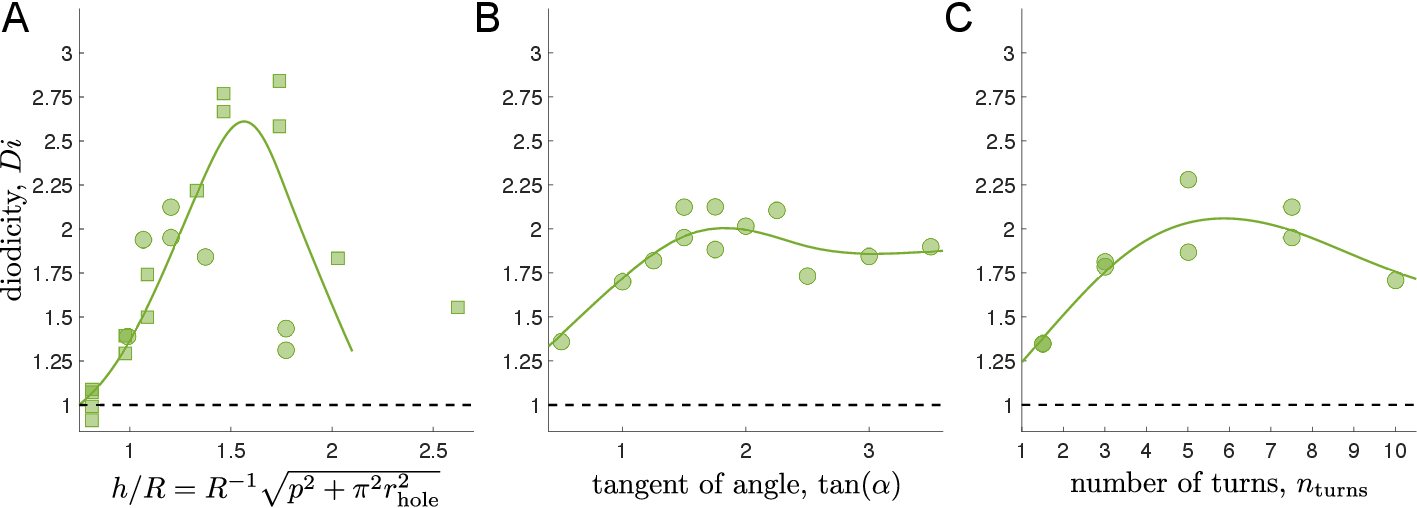}
\caption{
Dimensionless representation of \figRigid\ with fits.
(A) The dimensionless parameter $h/R=R^{-1}\sqrt{p^2+\pi^2 r_\text{hole}^2}$ collapses data from \figRigid A,B for pitch, $p$, and hole radius, \rhole\ (squares and circles, respectively).
The collapse is fit by the ratio of the splines in \figref{sfig:model}A (solid line).
The fit is not applied when the ratio of splines has large uncertainty, which occurs at small values of the effective length, $\ell$.
(B) and (C) Diodicities with respect to the tangent of the angle, tan($\alpha$), and the number of turns, \nturns\ (\figRigid C,D, respectively) fit by the ratio of corresponding splines in \figref{sfig:model}B,C (solid lines).
}
\label{sfig:Dicollapse}
\end{figure}

Next, in \figref{sfig:Dicollapse}A, we present a dimensionless version of this model. The data points in \figref{sfig:Dicollapse}A are values of diodicity versus a dimensionless helical length, $h/R$. The solid line is the ratio of splines from the model in \figref{sfig:model}.
The line fits the data well, except in regions where the data have large uncertainty, which arise when the helical lengths for helices oriented in the up and down directions are both close to 0, at values of $h/R>2$. 

We extend our model to propose that the effective lengths of rigid pipes in both the up and down orientations are a function of three parameters independently: $h$, $\alpha$, and \nturns.
Specifically, we hypothesize that effective lengths in the up direction are given by $\ell_{\uparrow}(h,\alpha,n_\text{turns})=H_{\uparrow}(h) \times A_{\uparrow}(\alpha)\times N_{\uparrow}(n_\text{turns})$, and effective lengths in the down direction are given by $\ell_{\downarrow}(h,\alpha,n_\text{turns})=H_{\downarrow}(h) \times A_{\downarrow}(\alpha)\times N_{\downarrow}(n_\text{turns})$, over the range of design parameters we tested. 
Using data that correspond to $h\approx$~12, we interpolate these functions using smoothing splines (\figref{sfig:model}).

To test this extended model, we introduce five new sets of data for effective length, $\ell$, versus the number of helical turns, \nturns.
The values $p$ and \rhole\ from these five new sets correspond to three new values of $h$ (roughly 8, 10.5, and 17.5).
For helices oriented downward, the new data fit our phenomenological model well for all new values of $h$  (\figref{sfig:model}D). For helices oriented upward, the new data fit our phenomenological model well for $h\approx$~10.5 and $h\approx$~17.5 (\figref{sfig:model}E).

Finally, in \figref{sfig:Dicollapse}, we present a dimensionless version of \figRigid.
The solid line is the ratio of $\ell_{\uparrow}$ to $\ell_{\downarrow}$, which is $\brk{H_{\uparrow} \times A_{\uparrow}\times N_{\uparrow}}/\brk{H_{\downarrow} \times A_{\downarrow}\times N_{\downarrow}}$ in the phenomenological model.
The line from that model fits the data well. 

\section{3D printing deformed pipes}
In the main text, we hypothesize that huge flow asymmetry in deformable, helical pipes is due to flow-induced deformation of the pipes' inner helical structure.
To test that hypothesis, we use the COMSOL finite-element solid-mechanics solver to find the deformation mode of the inner helix.
We make the assumption that the inner helix is subject to either uniform stress distribution in the flow direction or that it yields to its own weight under gravity.
Both scenarios result in similar deformations.

\begin{figure}
\centering
\includegraphics[width=15 cm]{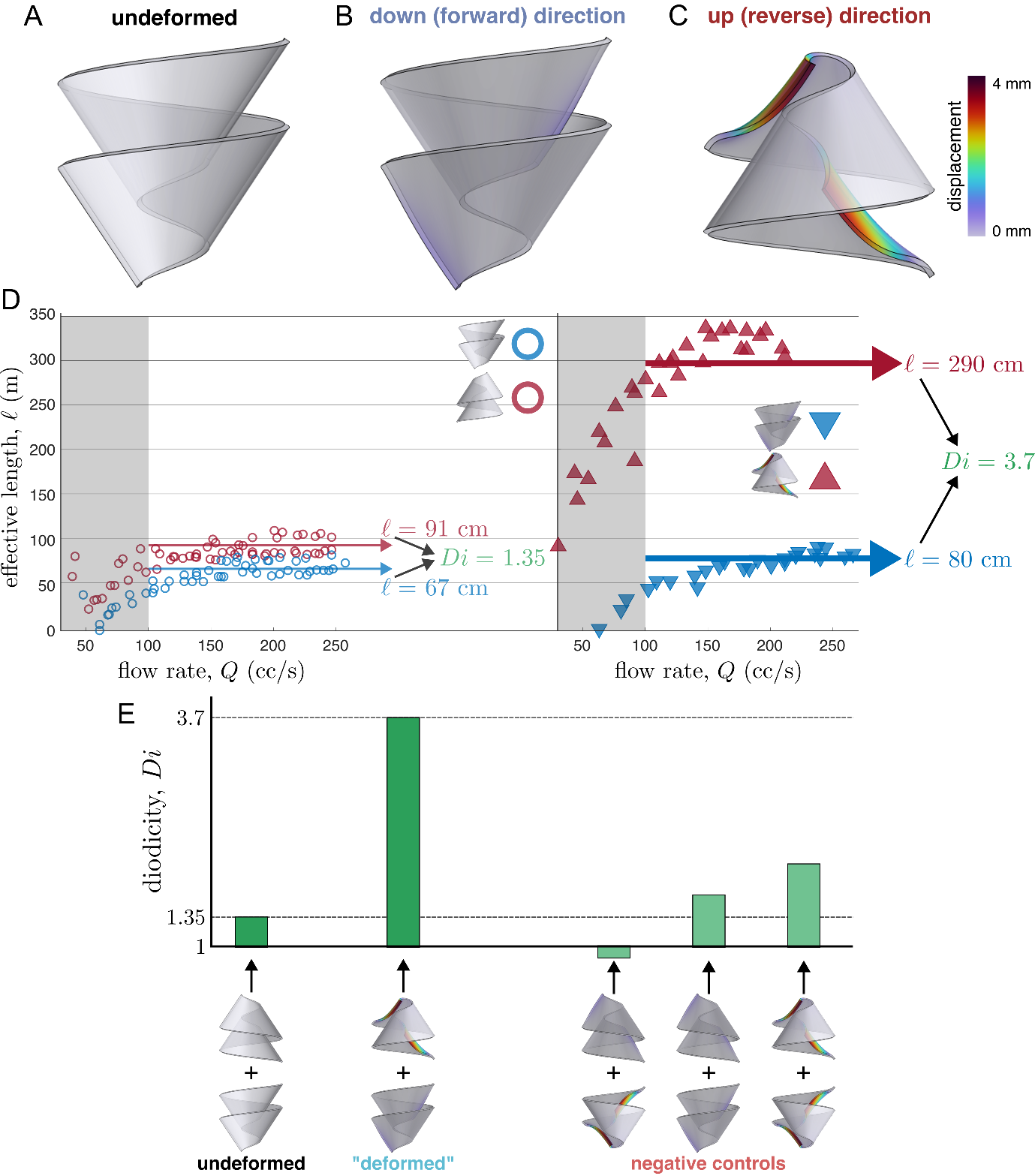}
\caption{
Deformation modes of soft helical pipes lead to large diodicities:
(A)~The inner helical structure of an undeformed pipe with $p=7.5$~mm, \rhole~=~3~mm, $\tan(\alpha)$~=~1.5, and \nturns~=~1.5.
(B) and (C)~Deformed structures of helices oriented in the down and up directions, respectively, calculated by COMSOL under a uniform load of 350~$\text{N/m}^2$.
The load was parallel to the symmetry axis (z-axis), pointing downward.
We 3D-printed rigid pipes containing the structures in Panels A, B, and C as the inner helices, using the same external tube dimensions and adaptors described in the main text.
We then used our flow apparatus to extract the effective length of each pipe in each orientation.
(D,~left~side) Experimentally measured effective lengths of pipes containing the original undeformed helices in panel A oriented in the down (blue) and up (red) directions. (D,~right~side) Experimentally measured effective lengths of pipes containing the deformed helices in Panel B (blue, downward triangle) and Panel C (red, upward triangle). The large diodicity of the deformed helix pair ($Di$~=~3.7) relative to the undeformed helix pair ($Di$~=~1.35) is primarily due to the large effective length of the deformed, upward helix, as in was in \figSoft.
(E)~Plot of the diodicities from Panel D and validation of results using three negative controls. The first control corresponds to deformed pipes assembled in opposite orientations than in Panel D. The second control corresponds to two orientations of the deformed helix in Panel B. The third corresponds to two orientations of the deformed helix in Panel C.
}
\label{sfig:deformation}
\end{figure}

When we increase the external load to 350~$\text{N/m}^2$, the helix's sheet nearly intersects with itself.
Therefore, we set 350~$\text{N/m}^2$ as the upper limit of the load.
From the resulting images in \figref{sfig:deformation}B and C, it is clear that: (1) most of the deformation is localized to the edges of the helix, and (2) most of the deformation occurs when the helix is oriented with its cone pointing upward; there is little deformation when the helix is oriented downward.
The first point is consistent with our experimental finding that diodicity depends only weakly on the number of helical turns, \nturns\ (\figSoft E).
The second point is consistent with our experimental finding that the effective length, $\ell$, of a helical pipe is nearly the same for soft and rigid pipes oriented downward (\figSoft B).

Next, we 3D-print two pairs of rigid pipes. The first pair consists of a rigid helical pipe with an inner helix equivalent to the undeformed shape  in \figref{sfig:deformation}A (parameterized by $p=7.5$~mm, \rhole~=~3~mm, $\tan(\alpha)=1.5$, and \nturns~=~1.5).
We experimentally test this pipe in our flow apparatus oriented either downward (blue circles) or upward (red circles). We calculate the corresponding effective lengths and take the ratio of these values to find the diodicity, which is $Di\approx1.35$ (\figref{sfig:deformation}D).

The second pair of pipes (shown on the right side of \figref{sfig:deformation}D) contains rigid inner helices equivalent to the COMSOL output of the deformed helices oriented downward (\figref{sfig:deformation}B, blue triangles) and upward (\figref{sfig:deformation}C, red triangles). We experimentally test this pair in our apparatus, calculate their corresponding $\ell$ values, and take the ratio of these values to find the diodicity, which is $Di\approx3.7$ (\figref{sfig:deformation}D). In other words, the diodicity due to deformed interior helices is roughly a factor of 3 larger than the diodicity due to undeformed helices (\figref{sfig:deformation}E, left side).

In addition, we test three pairs of helices as negative controls (\figref{sfig:deformation}E, right side). The first negative control shows that orientation matters: reversing the pair of helices in the "deformed" experiment in \figref{sfig:deformation} results in a diodicity close to 1 ($Di\approx0.85$). The second and third negative controls show that asymmetry in deformation in the up and down directions is required for large diodicities. For example, negative controls using rigid interior helices with only the shapes shown in \figref{sfig:deformation}B and \figref{sfig:deformation}C result in diodicities of only $Di\approx1.6$ and $Di\approx1.95$, respectively.

\section{Printing and testing replicas of shark intestines}
\begin{figure}
\centering
\includegraphics[width=12 cm]{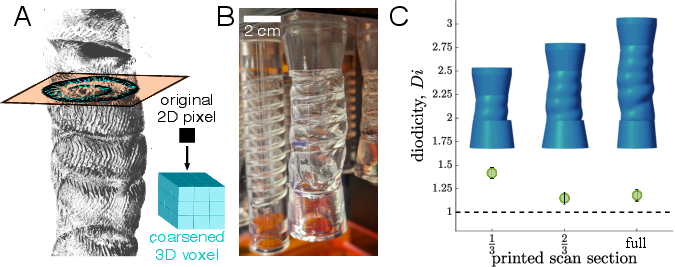}
\caption{
Measuring diodicity of structures based on computed tomography (CT) scans of shark intestines.
(A)~Data from a CT scan of \emph{Centroscyllium nigrum} \cite{leigh_lacmfish11156001_2020} was voxelated into a 3D structure.
The structure was then digitally coarsened, smoothed, and cut to the desired length (1/3, 2/3, or the full length). (B)~Conical adapters were then added to each end and the design was 3D printed. 
In the image, a print based on a shark intestine is in the foreground and a regular helical pipe is in the background at the left.
(C)~Diodicities of three printed versions of shark intestines, of increasing length.
Diodicity values are smaller than for most of the regular helical pipes in the main text but are comparable to values for traditional Tesla valves.
}
\label{sfig:CT_scan}
\end{figure}

To measure the diodicity of pipes based on computed tomography (CT) scans of shark spiral intestines,
we encode a scan of \emph{Centroscyllium nigrum}\cite{leigh_lacmfish11156001_2020} as a series of black-and-white bitmaps, each corresponding to a single slice along the z-axis.
These maps are voxelated into a 3D structure (voxel size $23 \times 23 \times 23~\text{\textmu m}^3$).
The raw structure is digitally coarsened via decimation, that is, every $4\times 4\times4$ super-voxel was added to the model if and only if it included at least one original voxel.
The coarsening is done in order to close holes in the model and remove some of the fine inner structures that were below the resolution of our 3D printer.
The coarsened model is upscaled by 300\% and rotated such that it was oriented along the z-axis.
The transformed model is imported into \emph{Autodesk Fusion 360} as a mesh, where it undergoes a repair process followed by re-meshing to remove any inaccuracies.
After the repair, the mesh is smoothed, resulting in smoother surfaces and improved model quality.
Then the smoothed model is sliced into overlapping segments of increasing length and conical adapters are fitted to the ends (\figref{sfig:CT_scan}A).

We 3D print the resulting structures using the rigid plastic (\figref{sfig:CT_scan}B) and attach them to tubes.
Diodicity measurements are performed (following the scheme for the rigid helical pipes).
The diodicities that result from shark intestine replicas are smaller than of most of the rigid helical pipes (\figref{sfig:CT_scan}C vs \figRigid).
However, the diodicities are comparable to values measured for traditional Tesla valves.

\begin{figure}
\centering
\includegraphics[width=17.5 cm]{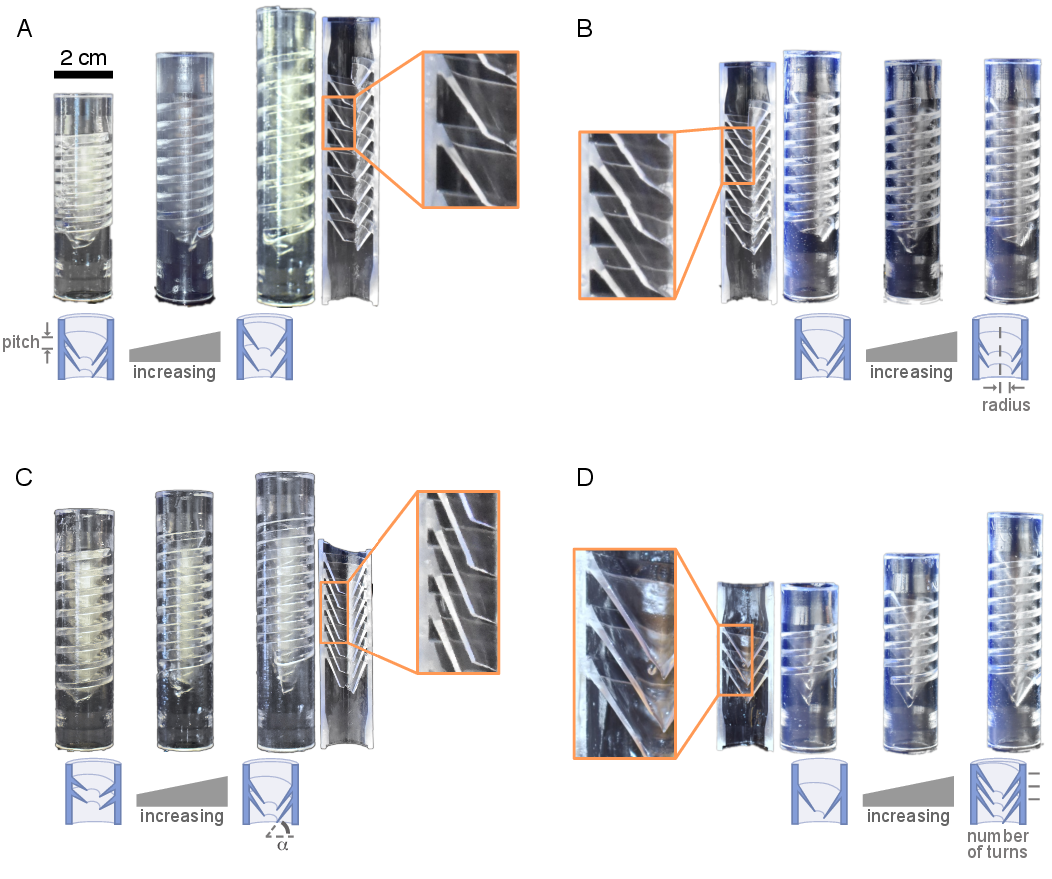}
\caption{
Sets of 3D-printed, rigid helical pipes in which a single parameter is varied.
(A)~Varying the pitch, $p$, for pipes with \rhole~=~3~mm, $\tan(\alpha)$~=~1.5, and \nturns~=~7.5.
(B)~Varying the hole radius, \rhole, for pipes with $p$~=~7.5~mm, $\tan(\alpha)$~=~1.5, and \nturns~=~7.5.
(C)~Varying the tilt angle, $\alpha$, for pipes with $p$~=~7.5~mm, \rhole~=~3~mm, and \nturns~=~7.5.
(D)~Varying the number of turns, \nturns, for pipes with $p$~=~7.5~mm, \rhole~=~3~mm, and $\tan(\alpha)$~=~1.5.
Each panel includes an additional print of a transverse section, revealing the inner helical structure, and an enlarged section (3x magnification, orange boxes) illustrating the printing quality.}
\label{sfig:3dprints}
\end{figure}

\begin{figure}
\centering
\includegraphics[width=11.4 cm]{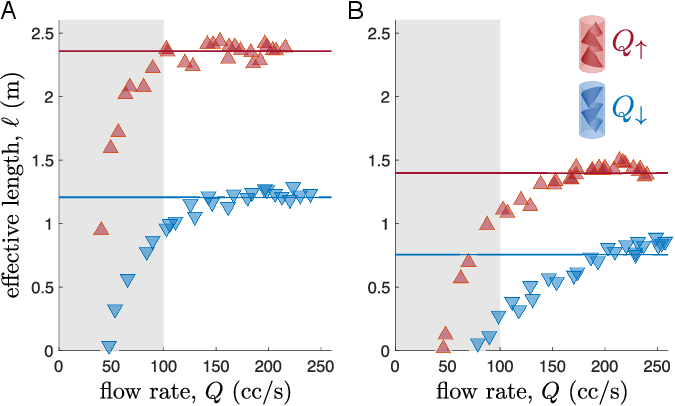}
\caption{
Estimates of effective length, $\ell$, of rigid, helical pipes with different pitches. (A)~Pitch~=~10~mm. (B)~Pitch~=~7.5~mm.
For both pipes, \rhole~=~3~mm, $\tan(\alpha)$~=~1.5, and \nturns~=~7.5.
Flow through pipes oriented in the down direction, \Qdn, are shown in blue, downward triangles, and flow through pipes in the up direction, \Qup, are shown in red, upward triangles.
The effective length is calculated using the calibration in \figref{sfig:Leff}.
Effective lengths rise to asymptotic values in the majority of our experiments, as in Panel A.
However, in a few cases, effective lengths do not reach an asymptotic value even at the maximum flow rate.
For each set of points, the characteristic effective length (horizontal line) is calculated by fitting a linear trend $Q_0 = a Q+b$ and converting the slope to an effective length.
For the fit, we exclude all points under the threshold of $Q=$100~cc/s (gray region), and use the 90\% uncertainty bounds as errors. 
Diodicity is calculated as the ratio between the effective lengths in the up and down directions.
}
\label{sfig:Leff_example}
\end{figure}

\end{document}